\documentclass[a4paper,11pt]{article}
\pdfoutput=1 

\usepackage{jheppub} 

\usepackage[T1]{fontenc} 
\usepackage{graphicx}
\usepackage{dcolumn}
\usepackage{bm}
\usepackage{subfigure}
\usepackage{amssymb}
\usepackage{amsmath}
\newcommand{\beq}{\begin{equation}}
\newcommand{\eeq}{\end{equation}}
\newcommand{\beqar}{\begin{eqnarray}}
\newcommand{\eeqar}{\end{eqnarray}}
\newcommand{\ds}{\displaystyle}
\newcommand{\bec}{\begin{center}}
\newcommand{\enc}{\end{center}}
\usepackage{color}

\title{\boldmath Classification of Equation of State in Relativistic 
       Heavy-Ion Collisions Using Deep Learning}

\author[a,b]{Yu.~Kvasiuk,}
\author[a,c,1]{E.~Zabrodin,\note{Corresponding author.}}
\author[a]{L.~Bravina,}
\author[d]{I.~Didur,}
\author[d]{M.~Frolov}

\affiliation[a]{
Department of Physics, University of Oslo,\\ PB 1048 Blindern,
Oslo N-0316, Norway}
\affiliation[b]{
Faculty of Physics, Taras Shevchenko National University of Kyiv,\\ 
Akademika Hlushkova Ave. 4, Kyiv UA-01033, Ukraine}
\affiliation[c]{
Skobeltsyn Institute of Nuclear Physics, Moscow State University,\\ 
Vorob'evy Gory, Moscow RU-119991, Russia}
\affiliation[d]{
DataRoot Labs,\\ Dmytrivska 80, Kyiv, Ukraine}

\emailAdd{yurii.kvasiuk@fys.uio.no}
\emailAdd{evgeny.zabrodin@fys.uio.no}
\emailAdd{larissa.bravina@fys.uio.no}
\emailAdd{didur@datarootlabs.com}
\emailAdd{frolov@datarootlabs.com}

\abstract{
Convolutional Neural Nets, which is a powerful method of Deep Learning,
is applied to classify equation of state of heavy-ion collision event 
generated within the UrQMD model. Event-by-event transverse momentum and 
azimuthal angle distributions of protons are used to train a classifier. 
An overall accuracy of classification of 98\% is reached for Au+Au 
events at $\sqrt{s_{NN}} = 11$~GeV. Performance of classifiers, trained 
on events at different colliding energies, is investigated. Obtained 
results indicate extensive possibilities of application of Deep Learning 
methods to other problems in physics of heavy-ion collisions.}	

\begin{document}
\maketitle
\flushbottom

\section{Introduction}
\label{sec:intro}

One of the main goals of physics of heavy-ion collisions is to explore 
properties of exotic state of matter, namely, hot, dense and 
hard-interacting baryonic matter. It can be recreated in laboratory by 
colliding heavy nuclei at relativistic energies. Lattice Quantum 
Chromodynamics (QCD) calculations indicate that a transition from 
quark-gluon plasma (QGP) to hadron gas is a smooth crossover at high 
energies and low baryon densities \cite{lattice}. It is widely believed 
that a first-order phase transition that ends up with tricritical point 
takes place within the energy range between $\sqrt{s} =  3$ and 
10~GeV, see, e.g., \cite{phdiagram} and references therein. Various past 
and ongoing experiments, such as beam energy scan (BES) and BES II at 
Relativistic Heavy Ion Collider (RHIC) \cite{besI,besII}, experiments at 
Super Proton Synchrotron (SPS) at CERN, are exploring collisions with 
gold and lead ion beams to find any peculiarities within the 
aforementioned energy range. However, neither the first-order phase 
transition, nor the tricritical point 
has been observed so far. Future experiments, such as Nuclotron-based 
Ion Collider fAcility (NICA) and Facility for Antiproton and Ion
Research (FAIR) are aiming to perform collisions at given energies with
higher luminosity, which give us a hope to see something new there. 
Difficulties in observing the phase transition originate from 
numerous factors. Some of them are an extremely short, approximately 
$10^{-24}$~fm/$c$, time of existence of the QGP phase, small number of 
particles in the system, high anisotropy of matter both in coordinate 
and in momentum space, etc. All valuable information registered by 
detectors is roughly several thousands of particles with corresponding 
energies and momenta. Therefore, it is immensely difficult to make any 
reasonable assumptions about the media they came from. 

Recent success of 
methods of Artificial Intelligence (AI), such as Machine Learning (ML) 
and Deep Learning (DL), in approximating highly obscure dependencies 
gives us a justifiable hope that they can benefit in heavy-ion physics 
objectives. Now these methods are widely employed in various problems 
that involve pattern recognition, for instance, image classification, 
natural language processing, and so on. Major advantage of such 
algorithms is that they do not need to be explicitly programmed. For 
example, AlphaGo \cite{alphago}, a program developed by Google Deep Mind 
to play a board game Go, managed to beat human world champion with the 
score 4:1. Played on 19x19 board, Go has roughly $2.08*10^{170}$ legal 
positions which makes it impossible for explicitly-written algorithm due 
to computational complexity. Moreover, a more advanced versions of the 
program, AlphaGo Zero and AlphaZero, were trained even without using the
human data or guidance, but only by playing between versions of 
themselves and reached even better results \cite{alphago2}. Recently, 
algorithms of DL have started to be applied to complex tasks in various 
fields of physics, namely, high-energy particle physics 
\cite{taudl,jetdl,partojetdl}, 
condensed matter physics \cite{manybodyann,vrgdl}, astroparticle physics 
\cite{cosmicraydl,astrophdl,exopldl}, and quantum information 
\cite{alphazq}. The work \cite{physconcdl} demonstrates how optimal 
representation variables, describing different states of physical 
systems, could be learned by neural nets. 
Generative Adversarial Nets, a generative Deep Learning models have 
started to be successfully applied for the event simulations 
\cite{jetgan,muongan}. Review of possibilities of application of ML and 
DL methods to high energy physics can be found elsewhere 
\cite{mlreview1,mlreview2,mlreview3}. Prominent results of application 
of DL to problems of high energy nuclear physics and heavy-ion 
collisions are described in \cite{eosdl,qcdtransitdl,spinoclumpml}.

Present work continues efforts of application of deep learning methods 
to identification of properties of equation of state (EoS). The paper
is organized as follows. Section \ref{sec:setup} introduces basic 
features of the microscopic ultrarelativistic quantum molecular dynamics
(UrQMD) model and parameters of hard and soft mean-field potentials used
in the calculations. It is outlined that the Deep Learning methods are
able to differentiate the calculations made with two different equations
of state on event-by-event basis compared to standard methods, which
need usually significant statistics of generated events. Principles of
the Convolutional Neural Network (CNN) algorithm are sketched in 
Sec.~\ref{sec:cnn}. Section~\ref{sec:results} presents main results of 
our study. Here we show also the ability of model, trained on 
particular collision energy, to make correct distinctions for the 
collisions at neighbor energies. Effectiveness of different training 
algorithms is discussed as well. Conclusions are drawn in 
Sec.~\ref{concl}.  

\section{Work Setup}
\label{sec:setup}

\subsection{Features of microscopic model}
\label{subseq:2.1}

The Ultra-relativistic Quantum Molecular Dynamics model
\cite{urqmd_1,urqmd_2} is a Monte-Carlo event generator designed for 
the description of $hh$, $hA$, and $A+A$
collisions in a broad energy range from hundred MeV up to several TeV.
The model contains 55 baryon and 32 meson states with masses up to 
2.25 GeV/$c^2$ as independent degrees of freedom, together with their 
antiparticles and explicit isospin-projected states. At energies 
below few GeV, the interaction dynamics of $hh$ or $A+A$ collisions can
be described via interactions between the hadrons and their excited 
states, resonances. At higher energies new processes of multiparticle 
production come into play. The UrQMD treats the production of new
particles via formation and fragmentation of specific colored objects,
strings. Strings are uniformly stretched between the quarks, diquarks
and their antistates with constant string tension $\kappa \approx 
1$\,GeV/fm. The excited strings are fragmenting into pieces via the
Schwinger mechanism of $q\bar{q}$-pair production, and the distribution
of newly produced hadrons is uniform in the rapidity space.
The model utilizes Hamiltonian dynamics of particle motions and 
incorporates particle interaction via geometric cross sections, taken 
from available experimental data or from quark models.

\subsection{Hard and soft EoS}
\label{subseq:2.2}

There is a possibility to switch-on potential interaction in the form 
of Yukawa, Coulomb, and Skyrme potentials 

\beq \ds
V = \sum_{i \neq j}\left(V^{Yuk}_0\frac{e^{-|\vec{r}_i-\vec{r}_j|/
    \gamma_{Y}}}{|\vec{r}_i- \vec{r}_j|}+\frac{Z_iZ_je^2}{|\vec{r}_i-
    \vec{r}_j|}\right) +t_1\rho^{int}_{j} 
    + t_{\gamma}(\gamma+1)^{-3/2}(\rho^{int}_{j})^{\gamma}\ ,
\label{eq1}
\eeq
where 
\beq \ds
\rho^{int}_{j} = \left( \frac{\alpha}{\pi} \right)^{3/2}e^{-\alpha r_j^2}
\label{eq2}
\eeq 

Here $\vec{r_i}$ is a coordinate vector of $i$-th particle, $Z_i$ is 
value of electric charge, $\rho^{int}_{j}$ is interaction density, 
$V^{Yuk}_0,\ \gamma_{Y},\ t_1,\ t_{\gamma}, \gamma$ are constants that 
define stiffness of EoS, see \cite{urqmd_1,urqmd_2,pot} for details. 
In present work we use potential with two different sets of parameters 
which can be associated with two separate equations of state, hard and
soft, of baryonic matter. Parameters of the hard and the soft potentials 
employed in the calculations are listed in Table~\ref{table1}.
\begin{table}[tbp]
\centering
\begin{tabular}{|c|c|c|}
\hline
Parameter      & Hard potential (EoS1)  & Soft potential (EoS2)   \\
\hline
   $\alpha$        &     0.25              &    0.25    \\
   $t_1$           &  -163                 &    -353    \\
   $t_{\gamma}$    &   125.93              &     304    \\
   $\gamma$        &     1.676             &    1.167   \\
   $V^{Yuk}_{0}$   &    -0.498             &    1.0038  \\
   $\gamma^{Yuk}$  &     1.4               &    1.4     \\
\hline
\end{tabular}
\caption{
Parameters of the hard and soft mean-field potentials used in the 
current version of the UrQMD.}
\label{table1}
\end{table}

\subsection{Example: directed flow}
\label{subseq:2.3}

It was shown earlier that EoS in UrQMD influences kinematic observables, 
such as directed flow, $v_1$, quite significantly \cite{flow}. Recall, 
that directed flow \cite{volflow} is an experimental observable that 
characterizes momentum space anisotropy of particle spectrum. It is 
defined as first coefficient of transverse flow Fourier decomposition 
\beqar \ds
E\frac{d^3 N}{d^3 p} &=& \frac{d^2 N}{\pi dp_T^2 dy} 
\Bigg\{ 1 + 2 \sum \limits_{n=1}^{\infty} 
v_n \cos{\left[ n (\phi - \Psi_n) \right] } \Bigg\}~,\\
\label{eq3}
v_n &=& \langle \cos{\left[ n (\phi - \Psi_n)\right] }\rangle~.
\label{eq4}
\eeqar
Here $p_T$ is particle transverse momentum, $y$ is rapidity, $\phi$ is
the azimuth between the $\vec{p_T}$ and the participant event plane, and 
$\Psi_n$ is the azimuthal angle of the event plane of $n$-th flow 
component.
Averaging in Eq.~(\ref{eq4}) is done over all particle of a given 
species from the single event and over all set of events.

\begin{figure}[tbh!]
\resizebox{\linewidth}{!}{
            \includegraphics[scale=0.60]{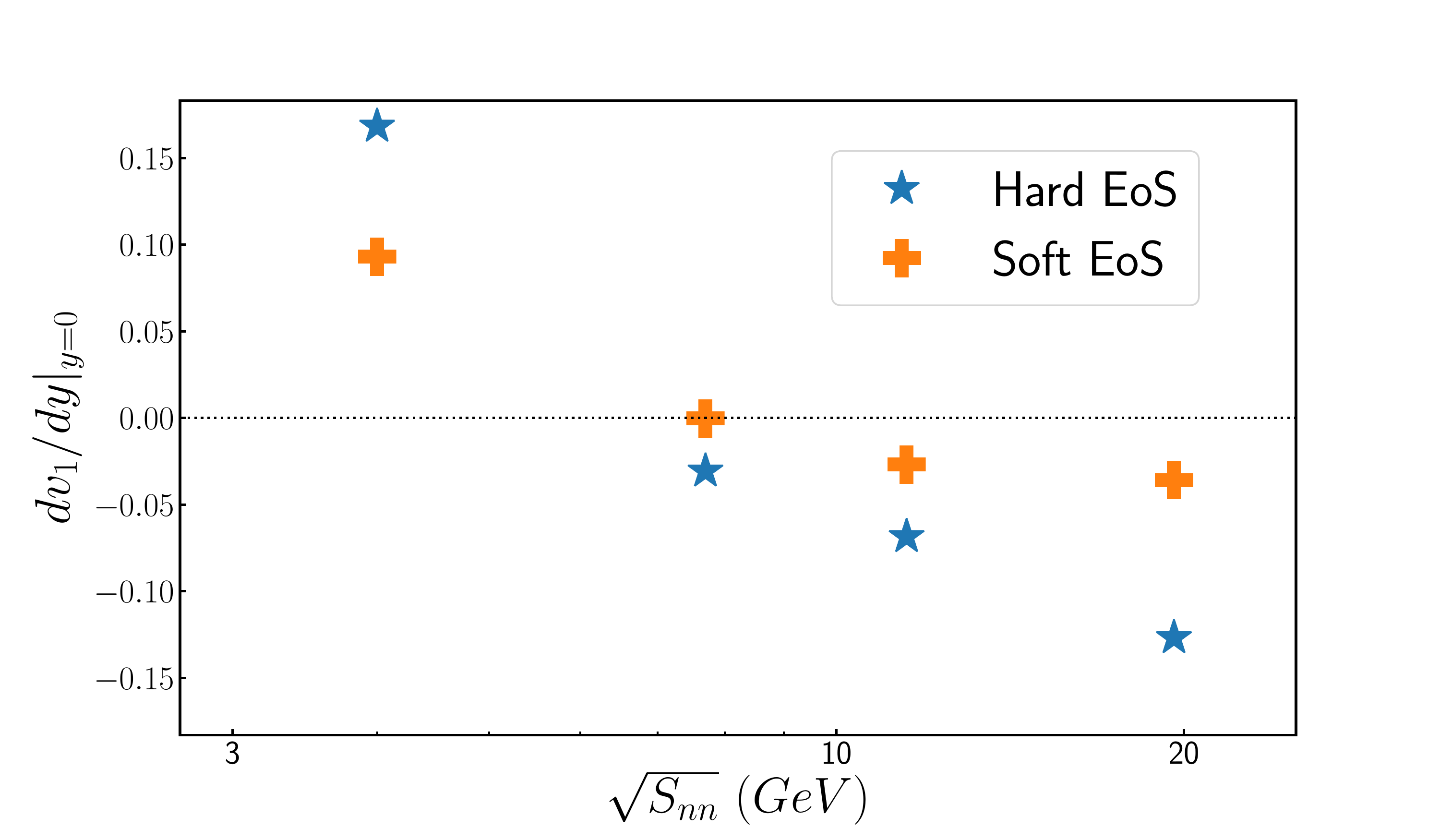}
}
\caption{
Slope of directed flow of protons coming from UrQMD minimum 
bias Au+Au collisions generated with hard (stars) and soft (crosses) 
EoS, respectively. Data are taken from \cite{flow}.}
\label{fig1}
\end{figure}

Figure \ref{fig1} shows the midrapidity slopes of directed flow of 
protons in minimum bias Au+Au collisions at energies $\sqrt{s} = 
4,\ 7.7,\ 11.5,\ 19.6$~GeV generated with two different EoS. One can 
see that the value of the slope is sensitive to the choice of EoS. 
However, to make the differences in $v_1$ slopes visible, about one 
million events for each data point were needed.

\subsection{Model setup}
\label{subseq:2.4}

It is worth mentioning that application of the Deep Learning methods 
permits one to differentiate between EoS on {\it event-by-event} basis. 
To check the efficiency of the method we generated 5000 Au+Au collisions 
of each kind at $\sqrt{s} = 11$~GeV with centrality 0-5\%. 
Distribution of proton number $n_{p} = n_{p}(p_T,\phi)$ was calculated 
for each separate event.
Transverse momentum and azimuthal angle were selected in ranges 
(0,1)~GeV/$c$ and (-$\pi$,$\pi$), respectively. Altogether, the training 
dataset is consisted of 10000 histograms with dimensions of $10 \times 
10$ bins and labeled "0" for hard potential and "1" for soft potential. 
All bin values were scaled to fit the range (0,1), so they represent 
probability density distribution. 

\section{Deep Convolutional Neural Network}
\label{sec:cnn}

\begin{figure}[tbh!]
\resizebox{\linewidth}{!}{
            \includegraphics[scale=0.60]{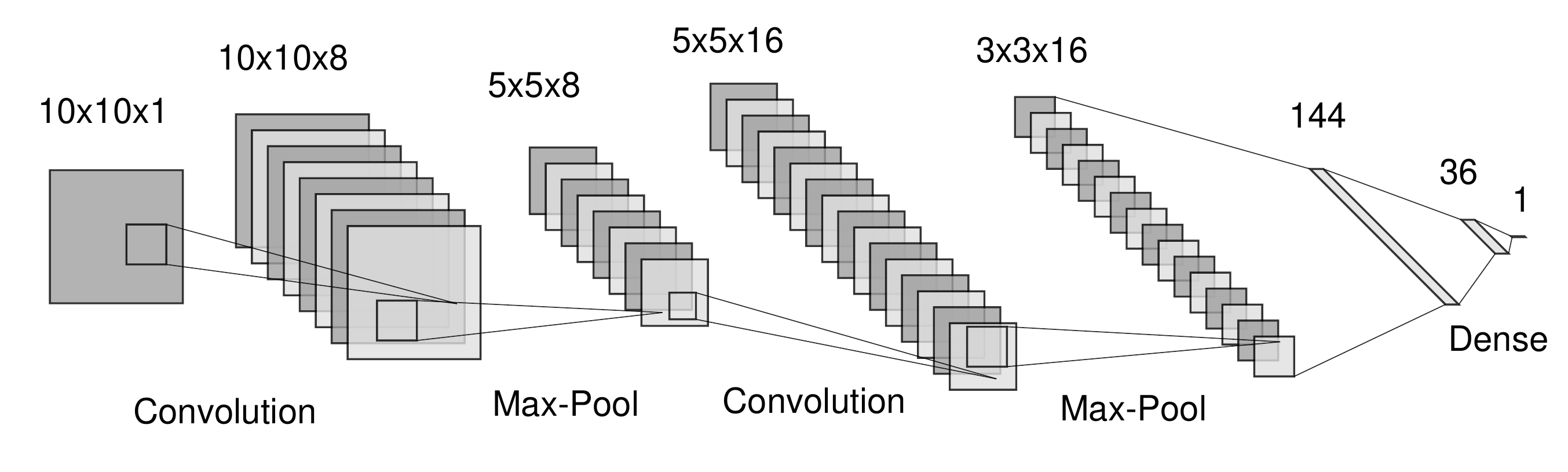}
}
\caption{Network architecture diagram.}
\label{fig2}
\end{figure}

Convolutional Neural Network (CNN) \cite{CNN} has proven to be a powerful 
deep learning algorithm for tasks dealing with graphical data, such as 
image recognition, classification, etc. It was inspired by the process 
of image processing in mammal's visual cortex. Similar to other deep 
learning algorithms, CNN consists of layers of artificial neurons. 
However, there are three main types of layers that are used in deep CNN. 
Convolutional layers are used to extract useful features. Pooling layers 
are responsible for reduction of parameters and fully-connected ones are 
used for making the final prediction. They map extracted features to the
desired output, which is either "1" or "0" in our case. Configuration of 
layers is determined by the complexity of the task. Network architecture 
employed here is described in Table~\ref{table2} and visualized in 
Fig.~\ref{fig2}, build with the help of online tool for CNN architecture 
visualization \footnote{An online tool for CNN architecture visualization:
http://alexlenail.me/NN-SVG/LeNet.html}.
\begin{table}[tbp]
\centering
\begin{tabular}{|c|c|c|}
\hline
	No.        & Layer          &Parameters             \\
\hline
	0.         & Input          & 1x10x10 tensor        \\
	1.         & Conv2D         & 8 kernels 3x3, ReLU   \\
	2.         & MaxPool2D      & 2x2 kernel, stride 2  \\
	3.         & Conv2D         & 2 kernels 3x3 , ReLU  \\
	4.         & MaxPool2D      & 3x3 kernel, stride 1  \\
	5.         & FullyConnected & 144 neurons, ReLU     \\
	6.         & FullyConnected & 36 neurons, ReLU      \\
	7.         & Sigmoid        &                       \\
\hline
\end{tabular}
\caption{Architecture of Convolutional Neural Network.}
\label{table2}
\end{table}
Binary cross-entropy, used as a loss function, reads
\beq \ds 
S = - w_n \left( y_n*\log{(x_n)} + (1 - y_n)*\log{(1-x_n)} \right)
\label{eq5}
\eeq
Here $w_n$ is a sample weight (set to 1), $x_n$ and $y_n$ are network 
outputs and true labels, respectively. Index $n$ runs over the batch. 
The network was trained with a help of stochastic gradient descent (SGD) 
method with learning rate 0.01 during 800 epochs. Dropout \cite{dropout} 
layers (p=0.3) were used before every fully-connected layer to reduce 
overfitting. Validation loss was recorded after every 10 epochs and 
model parameters were saved every time it decreased. Model parameters 
that yielded lowest loss at validation set were used for model 
evaluation afterwards. Eighty percent of the dataset was used for 
training and twenty percent for validation. Data acquisition and early 
preprocessing steps were done within the ROOT framework \cite{root}.
Further process was realized in Jupyter-lab \cite{jupyter-lab}, which
is an interactive Python notebook, with an extensive use of the Python 
libraries NumPy \cite{numpy_1,numpy_2}, Matplotlib \cite{matplotlib}, 
Pandas \cite{pandas}, and Scikit-learn \cite{skl}. The network was 
realized in PyTorch \cite{pytorch} framework.

\section{Results}
\label{sec:results}

\subsection{Classification accuracy for central Au+Au collisions at 
three different energies}
\label{subseq:4.1}

\begin{figure}
\resizebox{\linewidth}{!}{
            \includegraphics[scale=0.50]{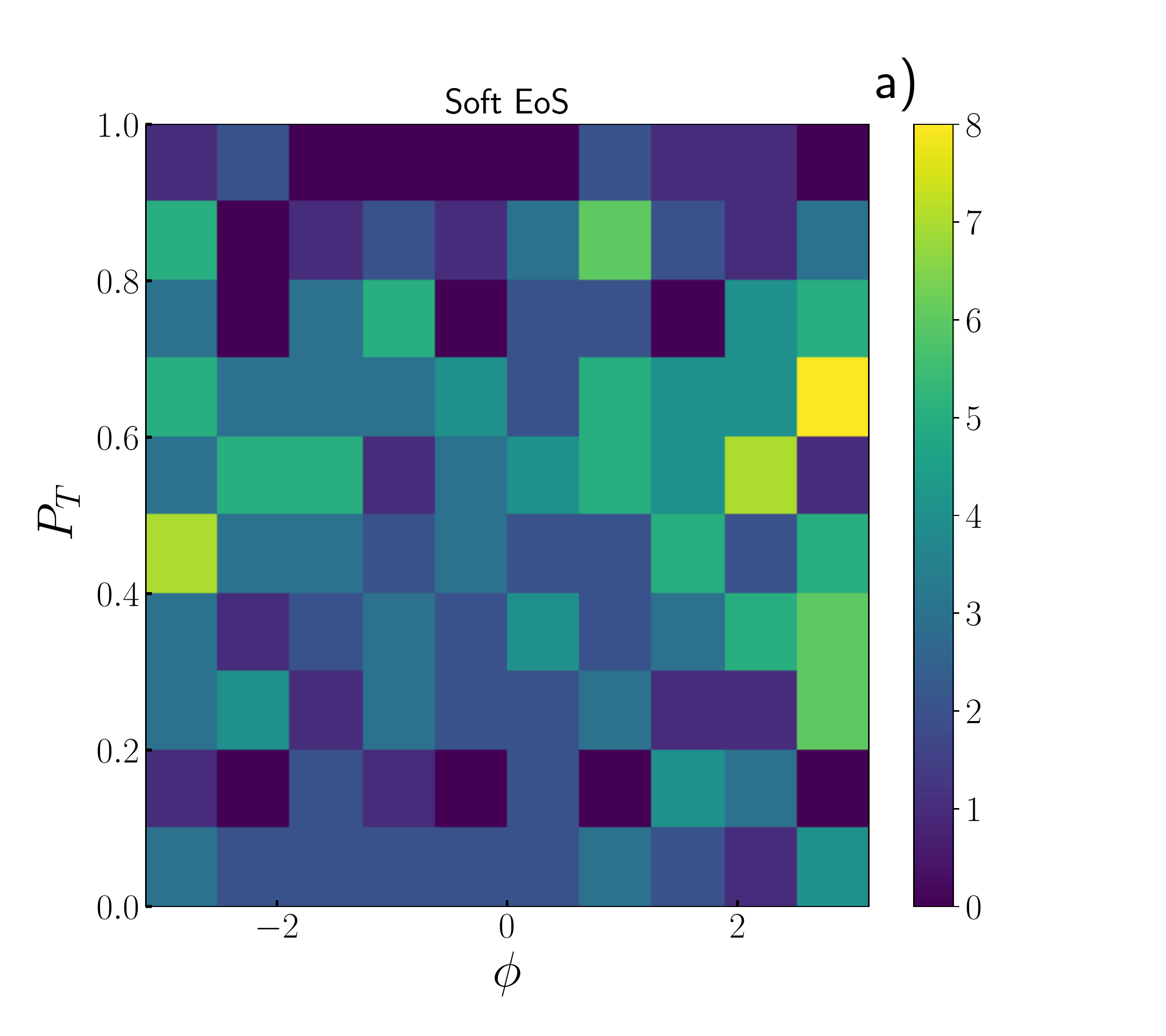}
            \includegraphics[scale=0.50]{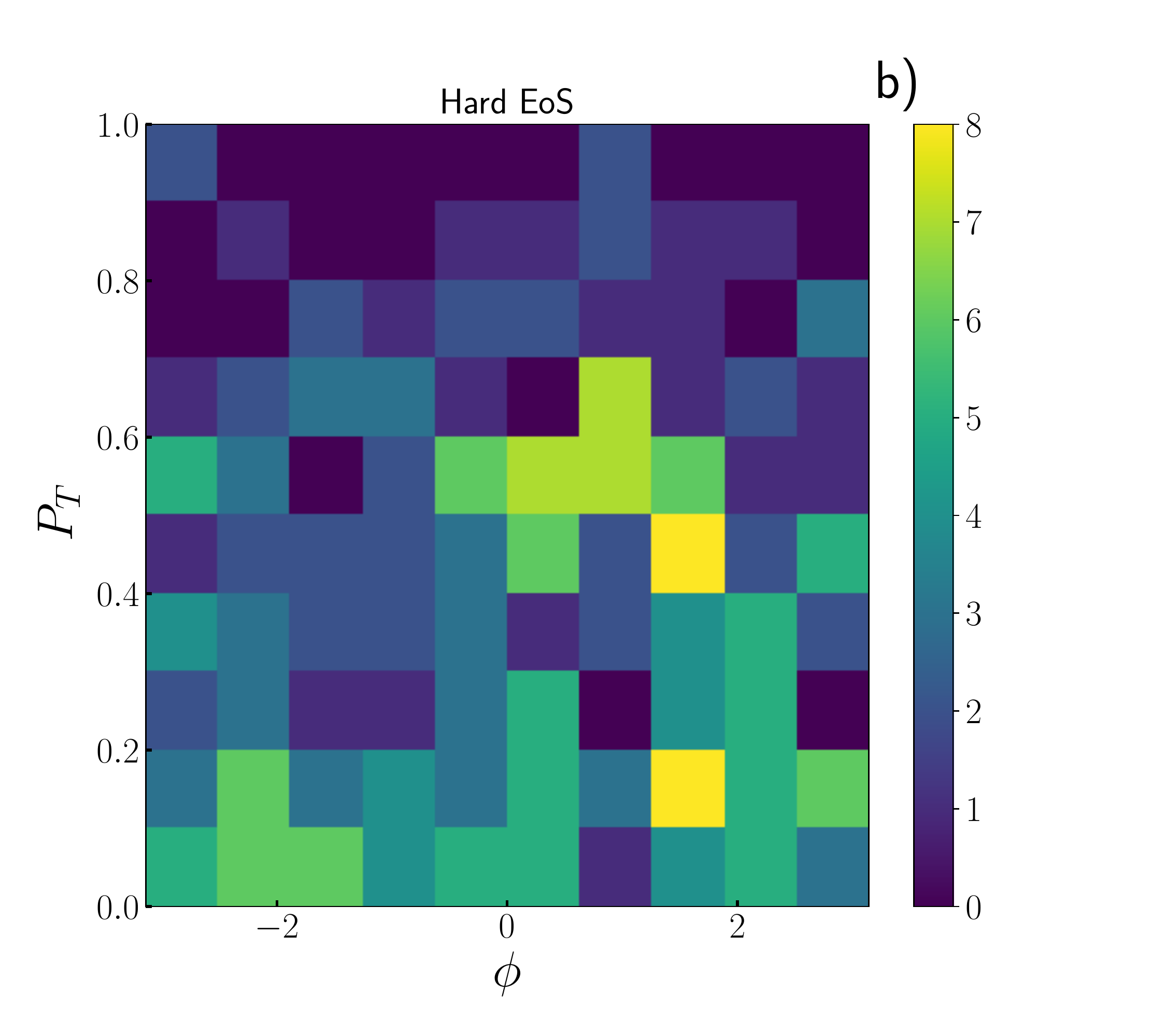}}
\caption{
Proton densities for (a) soft and (b) hard potentials from a single 
UrQMD generated Au+Au collision at $\sqrt{s} = 11$~GeV.}
\label{fig3}
\end{figure}

\begin{figure}
\resizebox{\linewidth}{!}{
            \includegraphics[scale=0.50]{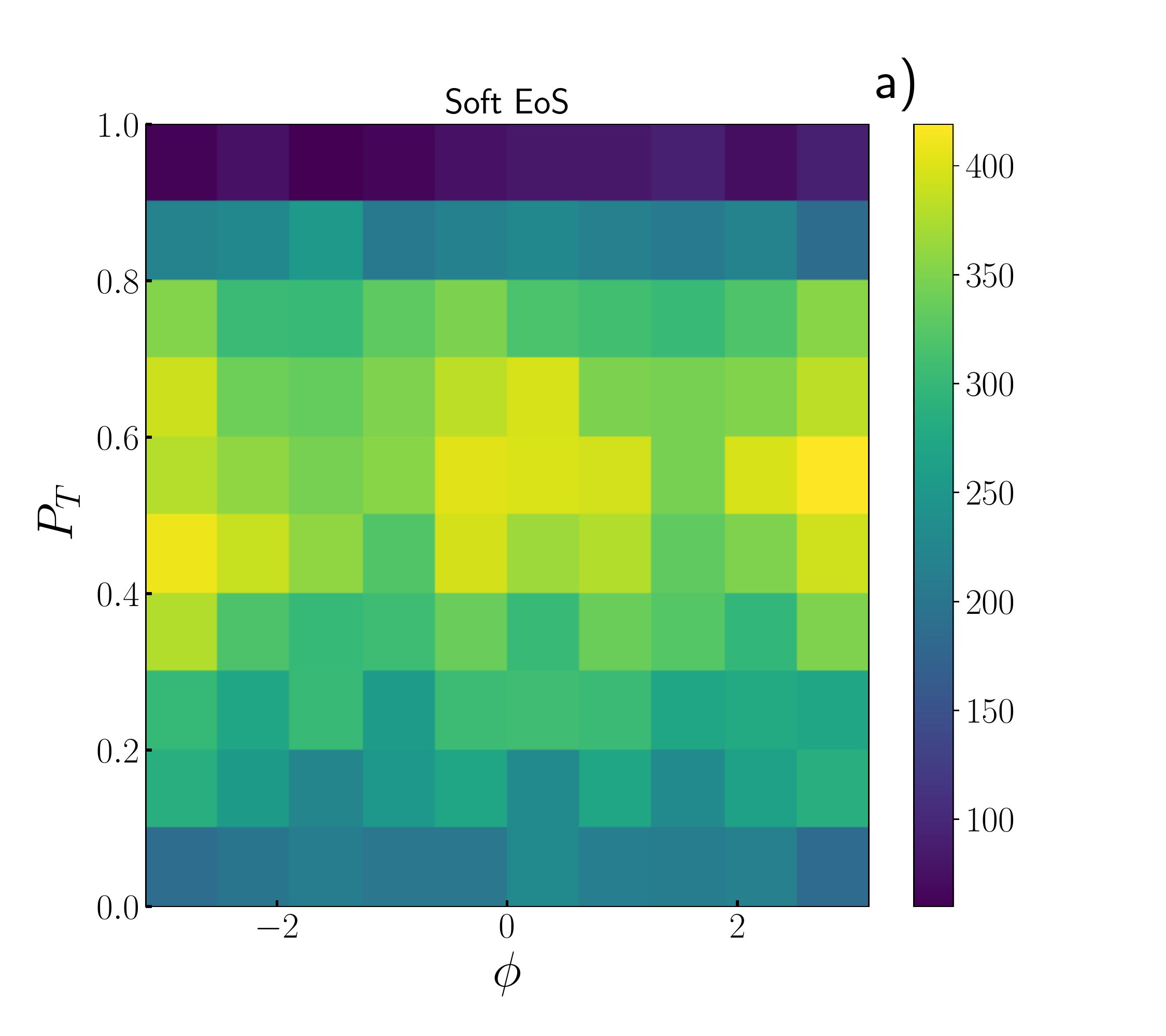}
            \includegraphics[scale=0.50]{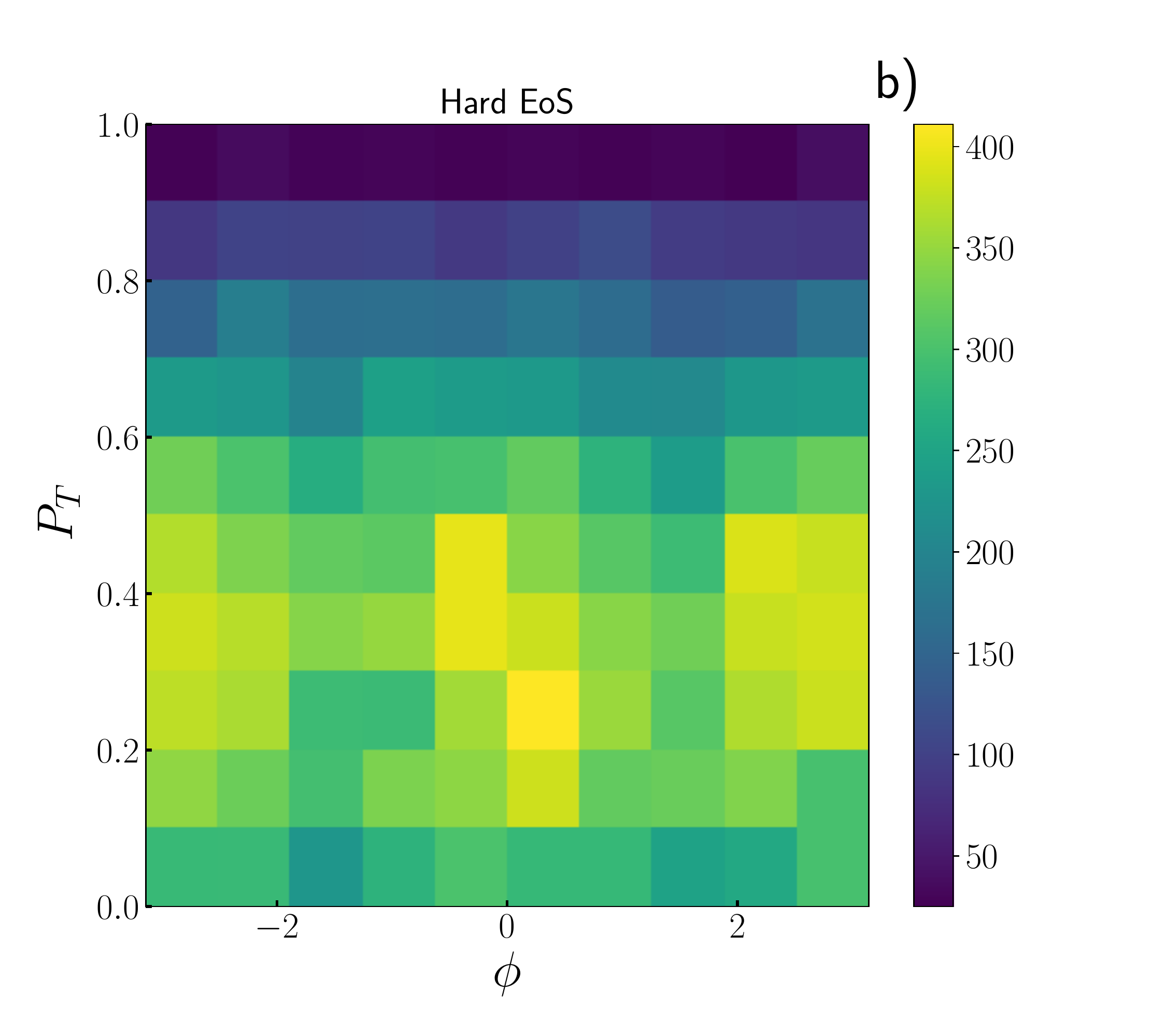}}
\caption{
The same as Fig.~\ref{fig3}, but for 100 Au+Au collisions.}
\label{fig4}
\end{figure}

Typical particle density profiles coming from events generated with 
both hard and soft potentials are shown in Fig.~\ref{fig3}.
We have to emphasize that density profiles from single event are 
strongly influenced by fluctuations and substantially differ on 
event-by-event basis. They are hardly distinguishable by naked eye, yet 
if we consider larger statistics, e.g., 100 events, the differences 
become more noticeable, as shown in Fig.~\ref{fig4}.

\begin{figure}
\resizebox{\linewidth}{!}{
            \includegraphics[scale=0.60]{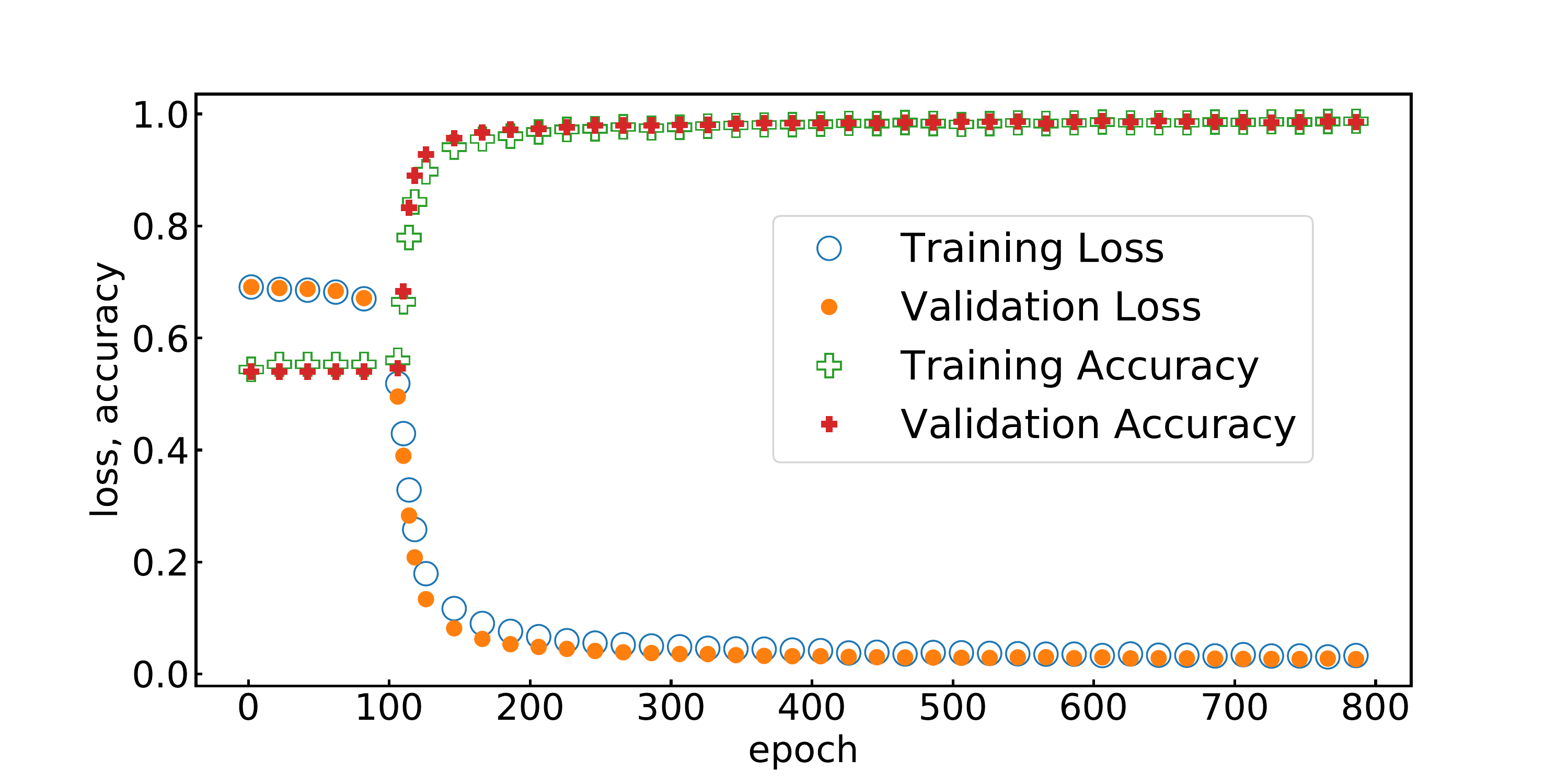}
}
\caption{
Training and validation losses and accuracies for UrQMD
generated Au+Au collisions at $\sqrt{s} = 11$~GeV}
\label{fig5}
\end{figure}

Overall classification accuracy of 98\% percent was achieved for both 
hard and soft potentials. It means that out of 100 items of every kind 
98 were classified correctly. Learning curves are depicted in 
Fig.~\ref{fig5}. One can see that after 100-th epoch both training and 
validation losses start to abruptly fall, whereas both accuracies 
rapidly grow. This circumstance indicates that a direction of minimum 
of loss function in the model parameter space is found. All curves 
enter the saturation region after 150-th epoch. The fact that training 
metrics coincides with validation metrics indicates that our model does 
not overfit and generalizes well.

Then, analogous setup was used for energies of $\sqrt{s} = 7$ 
and 14~GeV. For the last case overall accuracy of 98\% was reached, 
however, it drops to 94\% of accuracy for $\sqrt{s} = 7$~GeV collision 
energy within the proposed training scheme. It could be attributed to
low multiplicity of secondary particles, produced in a single event. 
Next, it is interesting to study the ability of a model, trained on one 
particular energy, to make predictions for another energy. 
These results are listed in Table~\ref{table3}. There are three 
trained models together with their predictions for Au+Au collisions at
three energies, $\sqrt{s} = 7,\ 11,$ and  14~GeV. Table~\ref{table3} 
reveals an interesting result that a classifier, trained on events at 
collision energy of $\sqrt{s} = 14$~GeV, shows good performance while
classifying the events at lower energies. The same is true for the
classifier trained at $\sqrt{s} = 11$~GeV. It could indicate on 
similarities in hadron spatial distributions in both reactions. 
In contrast, performance of the classifier trained on $\sqrt{s} = 7$~GeV 
is significantly worse. Possible explanation could be low particle 
yields in a single event at this energy which are not enough to make a 
distinction properly.
\begin{table}[tbp]
\centering
\begin{tabular}{|c|c|c|c|}
\hline
Training energies & \multicolumn{3}{c}{Efficiency (\%)}          \\
\cline{2-4}
$\sqrt{s}$ (GeV)  &    7     &       11      &      14           \\
\hline
    	 7        &    94    &       55      &      50           \\
        11        &    77    &       98      &      90           \\
        14        &    74    &       96      &      98           \\
\hline
\end{tabular}
\caption{Performance of models trained at different energies.}
\label{table3}
\end{table}
As it turned out, the smallest amount of data needed for accuracy better
than 96\% is only 100 images, i.e., only 1\% of the whole dataset, for 
energies $\sqrt{s} = 11$ and 14~GeV. For the energy of 7~GeV the best 
performance with the least amount of data used is 92\% with 400 training 
instances. However, training time is roughly 5 times longer compared to
4000 epochs for $\sqrt{s} = 11$~GeV. 

\subsection{Application of more efficient algorithm - Adam}
\label{subseq:4.2}

\begin{figure}[tbh!]
\resizebox{\linewidth}{!}{
       \includegraphics[width=0.6\textwidth]{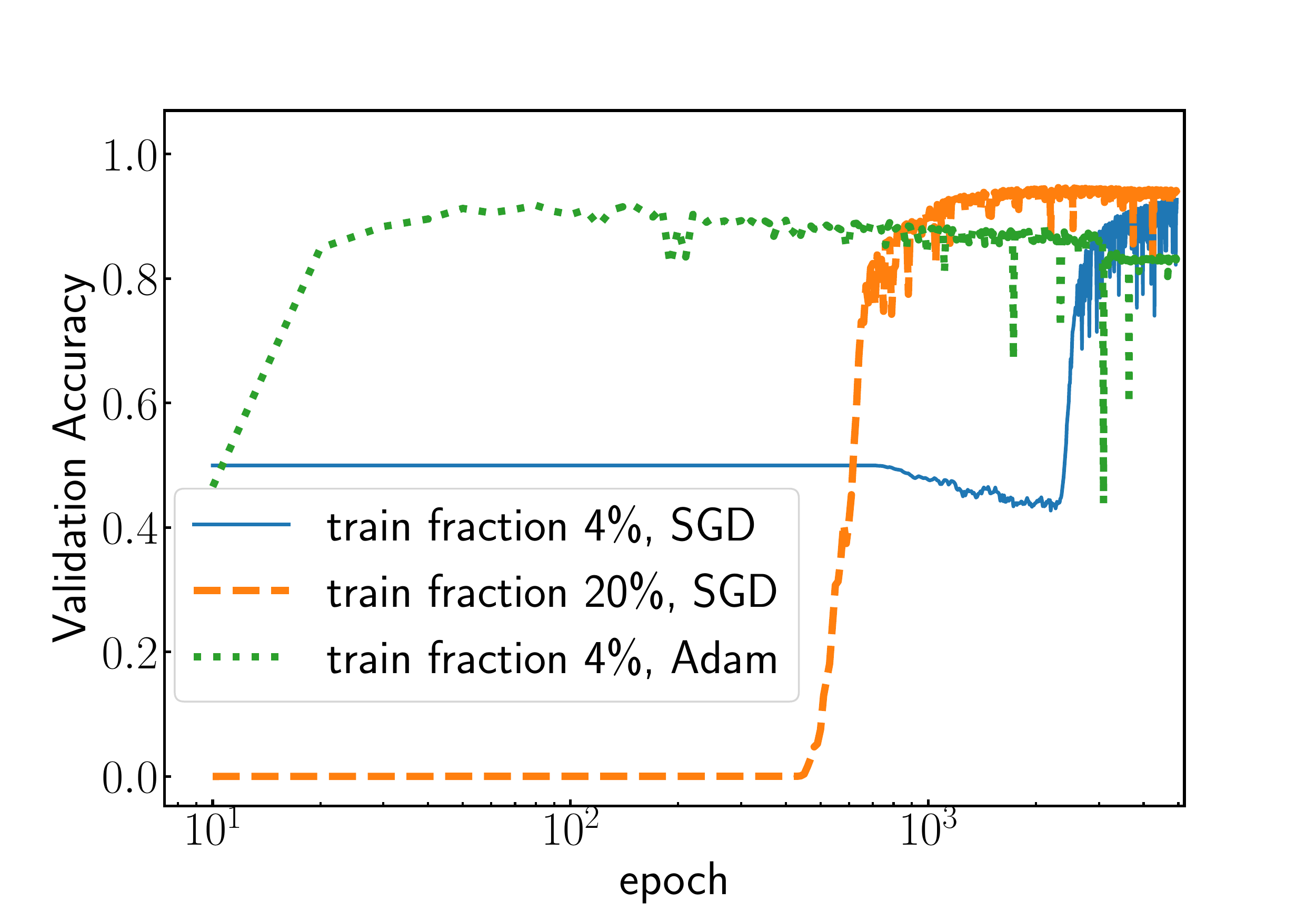}
}
\caption{
Validation loss during 5000 epochs of training 
for different training data fractions and optimizers in Au+Au central 
collisions generated within UrQMD at $\sqrt{s} = 7$~GeV. Solid and 
dashed curves indicate the results obtained by SGD method with
train fraction 4\% and 20\%, respectively. Dotted curve denote the 
results obtained by Adam method with 4\% train fraction.}
\label{fig6}
\end{figure}

\begin{figure}[tbh!]
\resizebox{\linewidth}{!}{
            \includegraphics[width=0.6\textwidth]{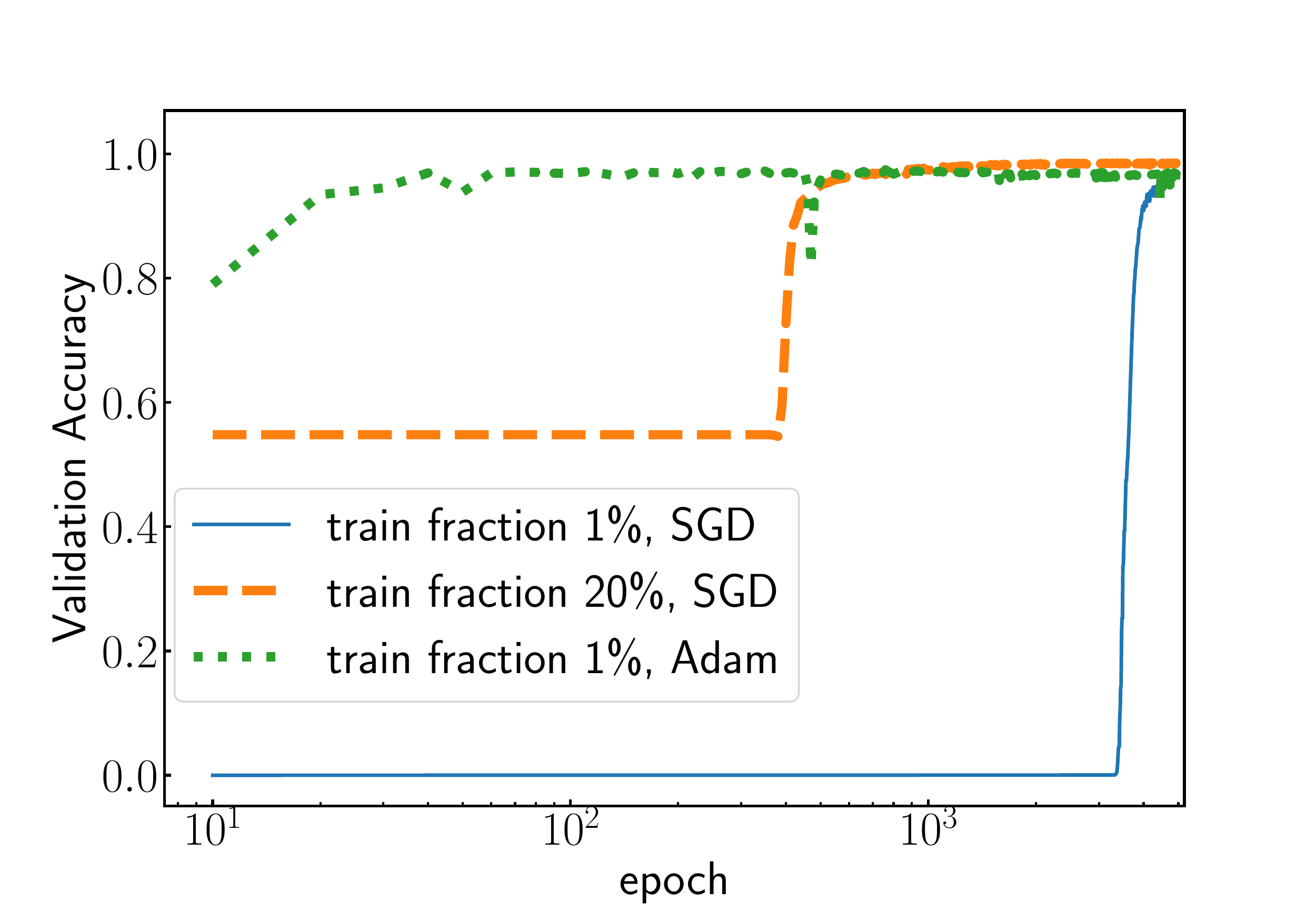}
}
\caption{
The same as Fig.~\ref{fig6} but for central Au+Au collisions at 
$\sqrt{s} = 11$~GeV.}
\label{fig7}
\end{figure}

\begin{figure}[tbh!]
\resizebox{\linewidth}{!}{
        \includegraphics[width=0.6\textwidth]{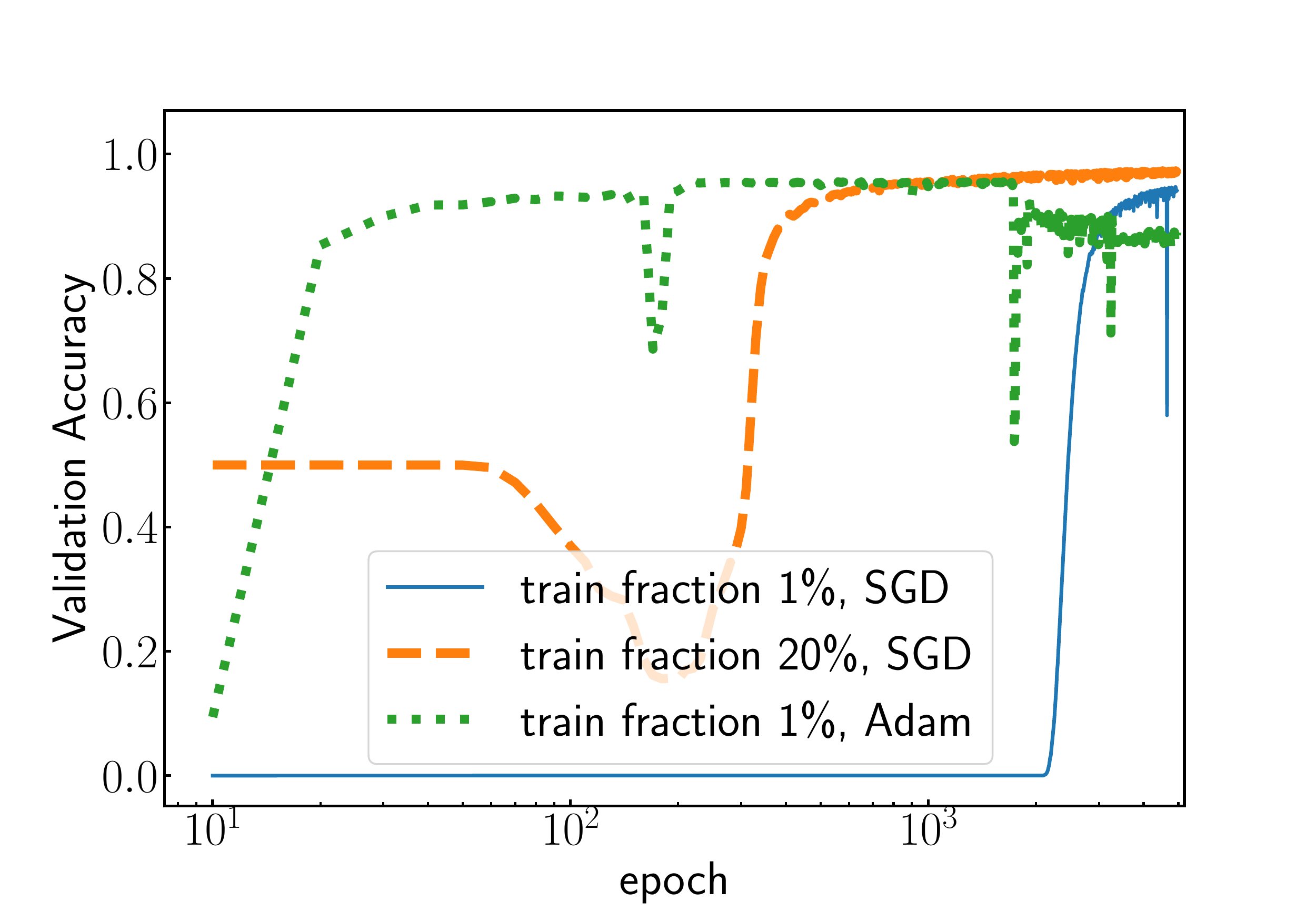}
}
\caption{
The same as Fig.~\ref{fig6} but for central Au+Au collisions at 
$\sqrt{s} = 14$~GeV.}
\label{fig8}
\end{figure}

It is possible to 
achieve very good results, namely, 92\% of accuracy on validation set 
for $\sqrt{s} = 7$~GeV and 96\% of accuracy on validation set for 
$\sqrt{s} = 11$ and 14~GeV, by using the Adam routine \cite{Adam}, which 
is a more advanced algorithm of loss minimization. Figure~\ref{fig6}, 
Fig.~\ref{fig7}, and Fig.~\ref{fig8} show overall validation accuracy 
during 5000 epochs of training for three energy data sets with different 
amount of training data and different optimizers. 

Results obtained for central Au+Au collisions at $\sqrt{s} = 7$~GeV are
displayed in Fig.~\ref{fig6}. To reach the desired accuracy above 90\%
by standard SGD method using just 4\% of generated statistics as a train
factor, one needs about 2500 epochs. This time can be shortened to
approximately 600 epochs if 20\% of events are used for the training.
In contrast, application of the Adam algorithm with 4\% train fraction 
needs only 30(!) epochs to reach the accuracy about 90\%. Peculiarity
in application of this method is the slow declining of the validation 
accuracy after 100 epochs, which means that the model starts to overfit.
Recall that training with low amounts of data is unstable with both 
training algorithms. More time is required by the model to find a path 
to the minimum of the loss. The problem can be cured by increase of the 
training statistics. It is important because the multiplicity of 
produced hadrons in heavy-ion collisions drops with decreasing collision
energy. For Au+Au collisions at $\sqrt{s} = 11$ and 14~GeV the 
multiplicities of secondaries are larger than that at $\sqrt{s}= 7$~GeV.

Here advantages of the Adam algorithm become more obvious, as shown in
Fig.~\ref{fig7} for the collisions at $\sqrt{s} = 11$~GeV. The standard 
SGD optimization method demonstrates sharp increase of the validation 
accuracy after 3200 epochs for the train fraction 1\%, and after 400 
epochs for the train fraction 20\%. Training with Adam algorithm 
converges much faster. We see a rapid increase of accuracy in EoS 
recognition with the train fraction 1\% after 30 epochs already. The 
same efficiency is obtained for central Au+Au collisions at $\sqrt{s} = 
14$~GeV, as depicted in Fig.~\ref{fig8}. Therefore, a classificator 
trained with the Adam optimization algorithm needs very limiting 
statistics for training within quite short time to discriminate between 
the hard and the soft equation of state with accuracy about 95\%.
  
\section{Conclusions and Discussion}
\label{concl}

Our main results can be summarized as follows.
Convolutional Neural Networks demonstrate excellent performance for 
equation of state classification in UrQMD generated heavy-ion collisions. 
Binary classification accuracy of 98\% is achieved for events at 
$\sqrt{s} = 11$ and 14~GeV and 94\% for $\sqrt{s} = 7$~GeV, using the 
data set of primitive experimental observables, such as particle 
transverse momentum and azimuthal angle. Moreover, it is possible to 
differentiate between equation of state using the model trained on 
different neighboring energies. 

It turned out that even a small fraction of data is enough to achieve 
satisfactory results. Training time depends strongly on data fraction 
used for the training, and on the employed algorithm. Our study shows
that to reach the top classification accuracy for central Au+Au
collisions at energies about 11~GeV and higher by the standard 
Stochastic Gradient Descent method, one needs more that 3000 epochs for 
training data of 1\% of the generated statistics, and more than 400 
epochs if the training fraction was increased to 20\%. Similar accuracy 
can be achieved by application of the stochastic optimization algorithm 
Adam trained at statistics of 1\% just after 30 epochs. Therefore, 
training Convolutional Neural Network classifier with the Adam 
algorithm permits one to reach high accuracy during very short training 
time with little amount of training data.

It is worth mentioning that the performance of trained CNN routine is
very robust against initial state fluctuations and final state
interactions, such as decays of resonances. However, similar to other
applications of DL technique to high energy physics 
\cite{eosdl,qcdtransitdl,spinoclumpml}, the
program has to decide between two discrete choices: calculations either
with soft or stiff equation of state. In other words, this is a
one-dimensional discrete space. In reality, the space of possible EoS
is continuous. The next step, therefore, should be calculations with a
combined EoS containing certain fraction $x$ of the soft EoS and
complementary fraction $1 - x$ of the stiff EoS. The model trained on
such data would definitely give us fractions of both EoS. It will take
more time because of the amount of generated data and longer training
procedure. But, as we show in our study, the application of sophisticated
algorithms, like Adam, makes solving this problem possible.

After that one can make a transition to multi-dimensional continuous
space. This will allow us to extract different parameters of the EoS
provided we can fix its functional form. If the limits of values of each
parameter, entering the EoS, are determined, we have to generate the
events for each set of parameters. Then, we train the model so that it
learns the correspondence between $(p_T,\phi)$-distributions of hadrons
and set of parameters. The whole procedure looks cumbersome but feasible.
This interesting problem needs further investigations.

\acknowledgments

Yu.K. acknowledges financial support of the Norwegian Centre for 
International Cooperation in Education (SIU) under Grant 
``CPEA-LT-2016/10094 - From Strong Interacting Matter to Dark Matter."
The work of L.B. and E.Z. was supported by Russian Foundation for Basic
Research (RFBR) under Grants No. 18-02-40084 and No. 18-02-40085,
and by the Norwegian Research Council (NFR) under Grant No. 255253/F50 -
``CERN Heavy Ion Theory."
Computer calculations were made at Abel (UiO,Oslo) and Govorun (JINR, 
Dubna) computer cluster facilities.

\end{document}